\documentclass[]{spie}  

\usepackage{amsmath,amsfonts,amssymb}
\usepackage{graphicx}
\usepackage{siunitx}
\usepackage{aas_macros}
\usepackage[colorlinks=true, allcolors=blue]{hyperref}

\title{Calibration of CMB Telescopes with PROTOCALC}

\author[a,b]{Gabriele Coppi}
\author[a]{Federico Astori}
\author[a]{Giulia Rancati Cattaneo}
\author[c]{Josquin Errand}
\author[d]{Rolando D\"{u}nner-Planella}
\author[a,b]{Federico Nati}
\author[a,b]{Mario Zannoni}
\affil[a]{Department of Physics, University of Milano-Bicocca, Piazza della Scienza 3, 20126 Milano, Italy}
\affil[b]{National Institute for Nuclear Physics (INFN), Sezione di Milano-Bicocca, Piazza della Scienza 3, 20126 Milano, Italy}
\affil[c]{Astroparticule et Cosmologie, Univ. Paris Cité, CNRS, Paris, France}
\affil[d]{Instituto de Astrof\'{i}sica and Centro de Astro-Ingenier\'{i}a, Facultad de F\'{i}sica, Pontificia Universidad Cat\'{o}lica de Chile, Av. Vicu\~{n}a Mackenna 4860, 7820436, Macul, Santiago, Chile}

\authorinfo{Further author information: (Send correspondence to Gabriele Coppi)\\E-mail: gabriele.coppi@unimib.it}

\begin{document} 
\maketitle

\begin{abstract}
Cosmic Microwave Background experiments need to measure polarization properties of the incoming radiation very accurately to achieve their scientific goals. As a result of that, it is necessary to properly characterize these instruments. However, there are not natural sources that can be used for this purpose. For this reason, we developed the \textbf{PROTOtype CALibrator for Cosmology}, PROTOCALC, which is a calibrator source designed for the \SI{90}{\giga\hertz} band of these telescopes. This source is purely polarized and the direction of the polarization vector is known with an accuracy better than \SI{0.1}{\degree}. This source flew for the first time in May 2022 showing promising result.
\end{abstract}


\section{INTRODUCTION}
\label{sec:intro}

Cosmic Microwave Background (CMB) polarization instruments need a very accurate control on the systematic errors introduced by absolute polarization orientation and polarized beam patterns, which limit the accuracy on multiple astrophysical signals such as the Inflationary Gravitational Waves and influences the reconstruction of the gravitational lensing effects on the CMB. Besides, an absolute calibration of the polarization angle of the CMB photons would enable the detection of signatures of parity-violating mechanisms in the early Universe, such as Cosmic Birefringence. Estimates for next-generation experiments such as Simons Observatory show that the accuracy required on the absolute polarization angle for measuring the tensor-to-scalar ratio $r$ with a delta lower than $2\cdot10^{-4}$ is around \SI{0.2}{\degree}\cite{abitbol2021}. However, this estimate is based on EB nulling techniques that are based on the assumption of the absence of parity-violation mechanisms. At the moment there are some evidence of the presence of Cosmic Birefringence\cite{Minami2020}. For this reason, it is necessary to have an independent calibrator for CMB telescopes. Unfortunately the best natural polarization calibrator, TAU-A, is not known well enough for CMB appplication\cite{Aumont2020}. Given this background we present the updated design and the realization of an artificial calibrator which should provide the polarization angle with an accuracy better than \SI{0.1}{\degree}. This system has been thought especially for the telescopes on the Atacama desert, such as Simons Observatory\cite{SO2019} and CLASS\cite{Class2014}, but can easily adapted to other sites.

\section{PROTOCALC}
\label{sec:protocalc}

PROTOCALC (PROTOtype CALibrator for Cosmology) is a project funded as a Marie-Curie Fellowship under the Horizon-2020 Program. The goal of the project is to develop a \SI{90}{\giga\hertz} polarization calibrator for CMB Telescopes with a polarization angle accuracy of \SI{0.1}{\degree}. We presented the device in \cite{Coppi2022}, but we significantly upgrade the system to achieve higher accuracy in terms of pointing and a longer flight time. 
The platform chosen for the upgraded version is still the same, so we use the DJI Matrice 600 Pro, and to keep the source stable we deploy a gimbal, in particular the DJI RONIN MX gimbal.

\begin{figure}
    \centering
    \begin{minipage}{0.45\textwidth}
        \centering
        \includegraphics[height=1\textwidth]{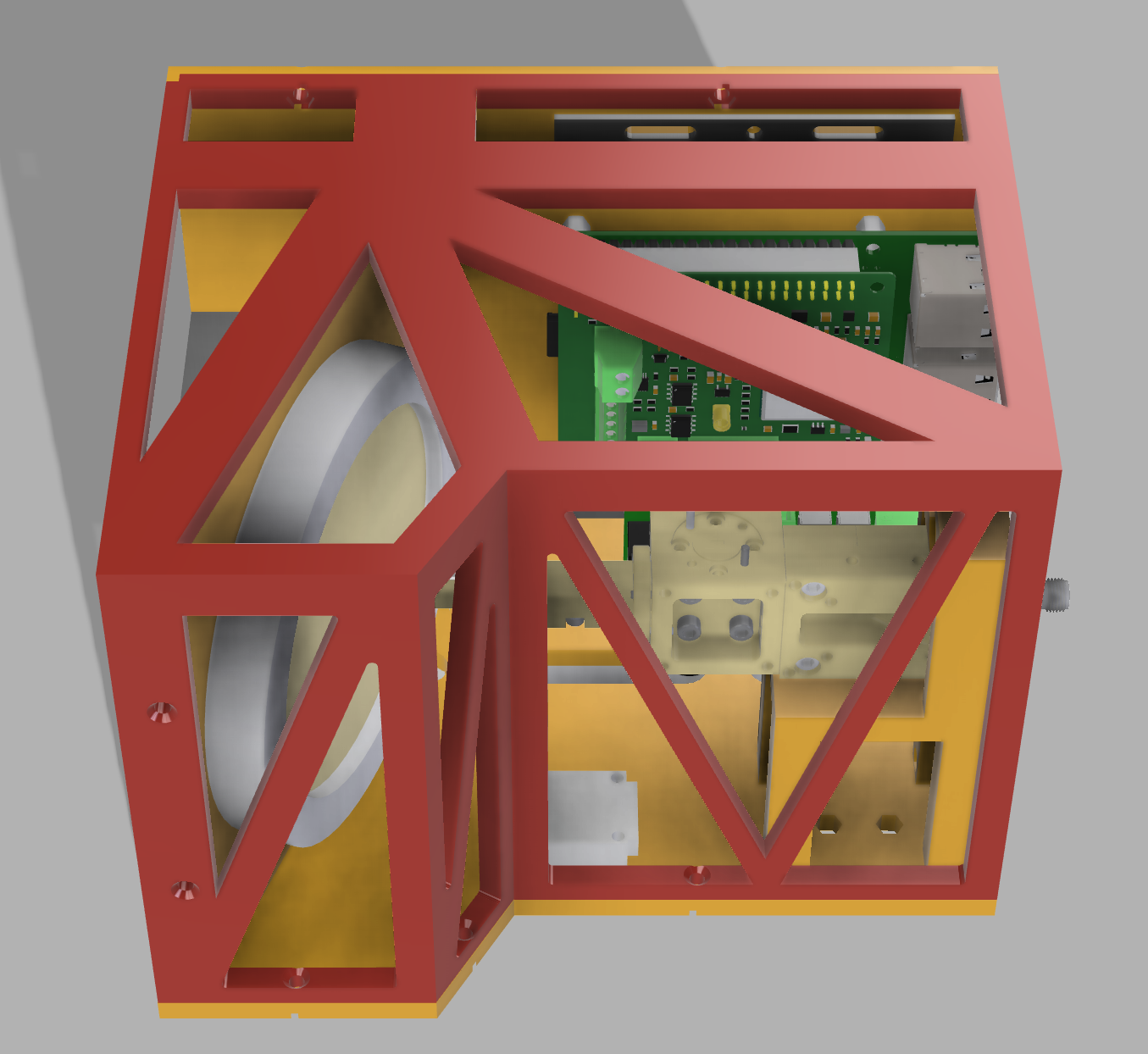}
        \caption{Rendering of the PROTOCALC updated payload.}
        \label{fig:render}
    \end{minipage}\hfill
    \begin{minipage}{0.45\textwidth}
        \centering
        \includegraphics[height=1\textwidth]{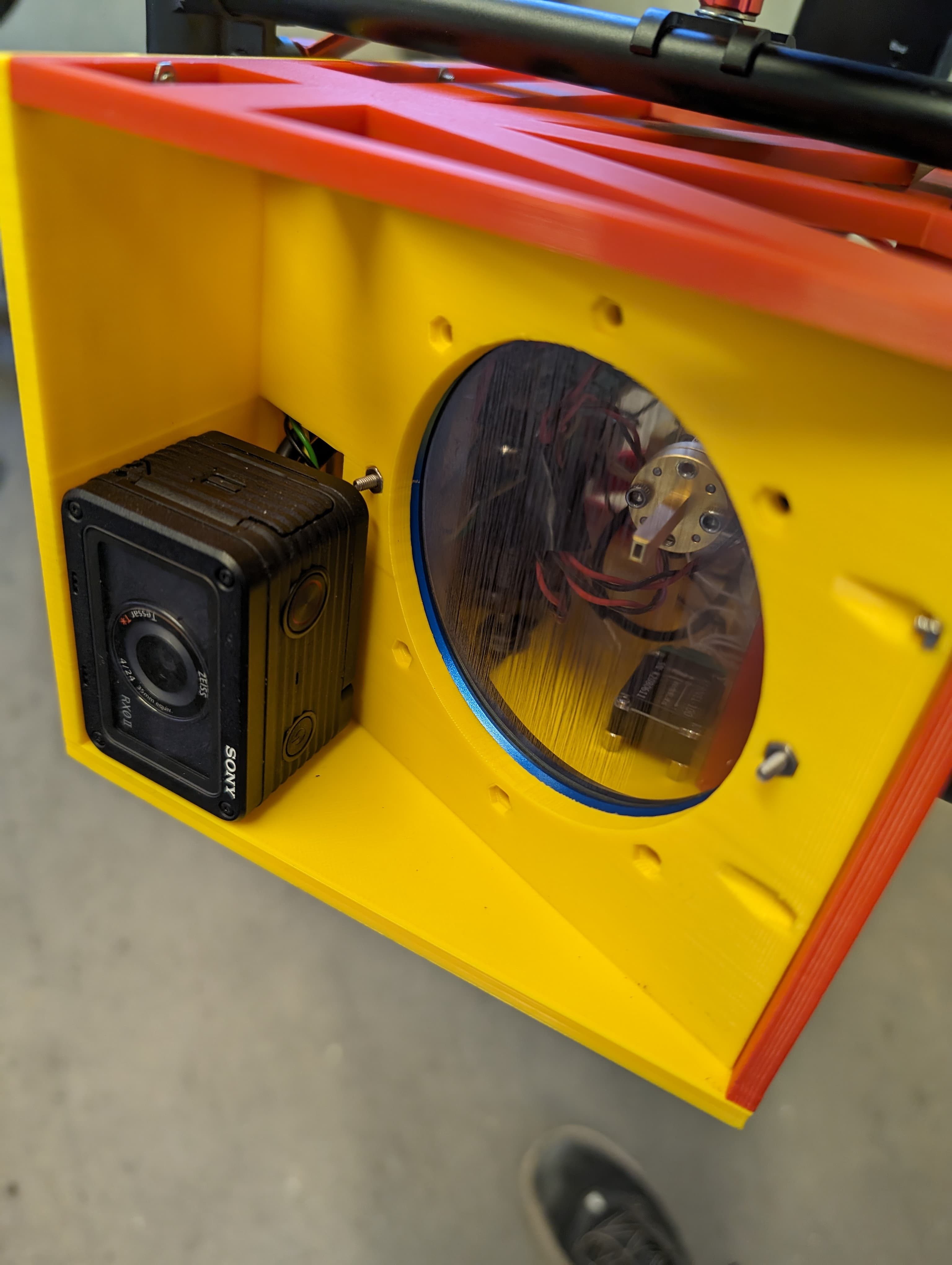}
        \caption{Payload assembled and mounted on the Gimbal.}
        \label{fig:photo}
    \end{minipage}
\end{figure}

\section{DESIGN}
\label{sec:design}

The philosophy design of the system is still the same, however we integrated several improvements based on the lesson learned. In particular, we focused especially in reducing the weight and optimizing the components on board. This means that we increase the number of sensors to compute the attitude and achieve the requirements. The volume constraints are still the same, but we found them not limiting our ability to develop a system that it is suitable for our applications. 

\subsection{New Mechanical Design}
\label{subsec:mechanical_design}

Compared to the previous design, we have updated the payload reducing the weight and changing material. We implemented a new 3D printed payload made of PLA, that reduces the total weight of about $25\%$. This weight reduction helps in increasing the flight time, thus calibration time. The total footprint of the new 3D printed payload is the same of the previous Aluminum Version, so a box of \qtyproduct[product-units=power]{16 x 16 x 13}{\cm}. The payload is composed by two parts, this allow an easy access to all the components in the system and easy to assemble. The parts are designed so that the top part can be slotted in ensuring a stable contact between the two parts. Additionally, pass-through holes allow to use screws and nuts to increase the rigidity of the system. 
To compensate for a lower thermal conductivity material to dissipate the heat, we employed a design with multiple holes to increase the radiative exchange with external and increase convective cooling.
As for the previous version the system will be mounted on a gimbal and the inteface between the payload and the gimbal is a metal dovetail. We decided to use this instead of 3D printing together with the payload for safety reason. Indeed, when we tight the payload on the gimbal, we noticed some tears in a prototype version of a 3D printed dovetail. 
Nevertheless, of the use of 3D printed PLA instead of machined Aluminum, the accuracy of the final printed payload is high enough for our application. In particular, the alignment calibration described in \cite{Coppi2022, Carrero2021} is replicated for the payload showing similar results.
The render of the payload and the the realized version are presented in Figure \ref{fig:render} and \ref{fig:photo}, respectively.

\subsection{Mechanical and Thermal Simulation}
\label{subsec:simulations}

\begin{figure}
    \centering
    \includegraphics[scale=0.35]{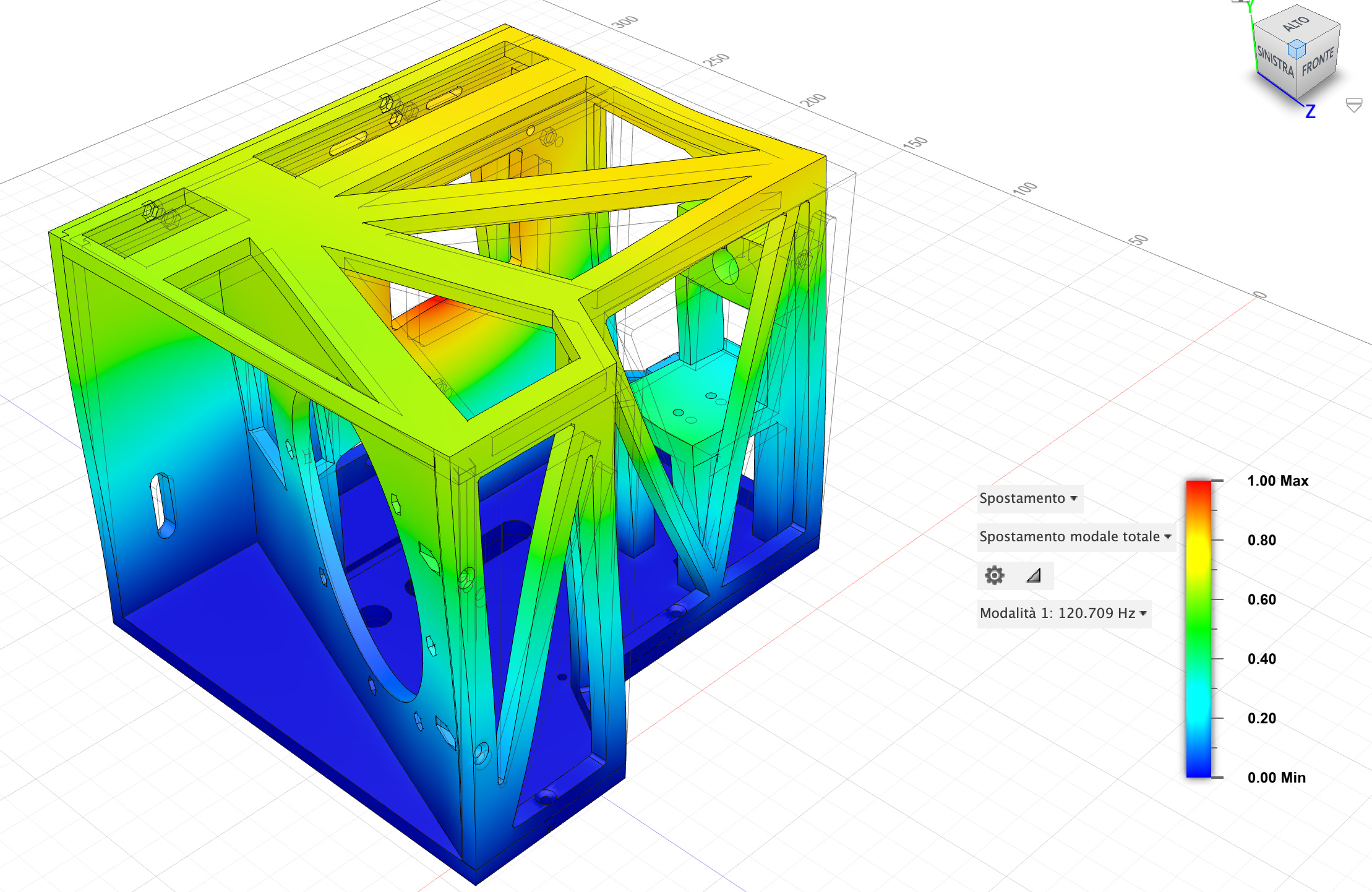}
    \caption{Simulation result from the Modal Analysis using Fusion 360.}
    \label{fig:simulation}
\end{figure}

The new payload design poses significant challenges compared to the aluminum one. Indeed, PLA has a lower tensile strength than Al-6061 and also the thermal properties are completely different. The boundary conditions of the problem are the same for both payloads, so we need to consider an operational temperature of \SI{0}{C} and the payload flying at $\sim$\SI{5200}{\meter} where there the atmospheric pressure is only half compared to the sea level. Compared to the previous version, we used a RPi4 but given code optimization and OS optimization the power consumption is similar to the RPi3 used in the previous flights. Given the reading of the inclinometer in the previous flights, we noted that we needed to pay attention particularly to study the resonance frequency of the structure. While PLA as a 3D printed material is orthotropic, the full structure can be considered isotropic. Indeed, we printed the different components using different orientation so that the full assembled payload will have similar properties along all axis. 
The result of the simulation are presented in Figure \ref{fig:simulation}. The simulations show that the first resonance frequency is to be found at \SI{120}{\hertz} which is high enough based on the previous flights experience. Thermal simulations show no particular critical points and the displacement between the camera and the polarizing grid is still negligible, $\le$ \SI{15}{\micro\meter} (a bit higher than aluminum but still acceptable for the scope of the project). 

\subsection{Software and Pointing Updates}
\label{subsec:software}

A big update compared to the previous version is the use of a Raspberry Pi 4. Nevertheless of its higher power consumption, the RPi4 has multiple UARTs compared to the single one available on the RPi3. The flight code has been completely re-written to include also a python version of the camera control system. The camera control software initialize the camera and set the time of the camera based on the internal clock of the RPi. The accuracy of the time initialization is between $2$ and $3$ $ms$ which is significantly lower than the distance between the frames of the video, which is recorded at $30$ fps. Additionally, we also record in the Log File when we send the command to start recording. The time on the RPi is regulated by the GPS with PPS and by a an external Real Time Clock. The control software is still highly threaded with a dedicated thread to each sensor.

In the previous generation of PROTOCALC, we had multiple sensors stacked on top of the RPi4 GPIO which creates a full payload. In this new updated version, we created a single board that behaves as a single RPi HAT with multiple sensors on top, as well as all the ports required to communicate with the Valon and the external inclinometer. The boards mounts a GPS, an ADC, a DAC, a RTC and a series of auxiliary sensor such as an IMU, magnetometer and a barometer. These auxiliary sensors were not used in the 2024 campaign since they still need to be calibrated, but future campaign will include the use of these sensor for attitude calculation. Attitude calculation will also include the gyroscope and accelerometer data coming from the video. Indeed, the Sony RX0 MII records these data as metadata and these can be extracted using available open-source software like \emph{telemetry-parser}\footnote{https://github.com/AdrianEddy/telemetry-parser}. The DAC is used to control a Virginia Diodes multiplier in Y-band for future missions. This multiplier has been provided by APC and CNRS and it has not been used yet, since in April 2024 Simons Observatory did not have yet deployed the \SI{220}{\giga\hertz} and \SI{280}{\giga\hertz}.

Regarding the passage to the new RPi4, we expected a bit higher power consumption but based on the laboratory tests and flight data, we did not see a significant difference. To decrease the power consumption, we set the RPi only in SSH mode, so no graphic mode output, which reduces the load on the GPU and so the power consumption. Additionally, the code optimization reduced the CPU utilization too. It is important to notice that we are not currently limited in our calibration campaign by the power consumption of the payload, but by the flight time of the drone. Indeed, the power source for the payload is not coming from the drone, but it comes from the gimbal battery which is separate from the drone batteries. Moreover, we re-disegned completely the power distribution breakout board to include high-efficiency components to convert the \SI{13}{\volt} given from the gimbal, to the \SI{5}{\volt} or \SI{8}{\volt} required by the different components on board.

\section{Flights Data}
\label{sec:first_flight}

The PROTOCALC source was flown three times between 2022 and 2024, however only the last flight campaign was performed using the plastic 3D printed payload. The first two campaigns focused only on CLASS while the last one focused on both CLASS and Simons Observatory Small aperture Telescopes. 

\begin{figure}
    \centering
        \includegraphics[scale=0.15]{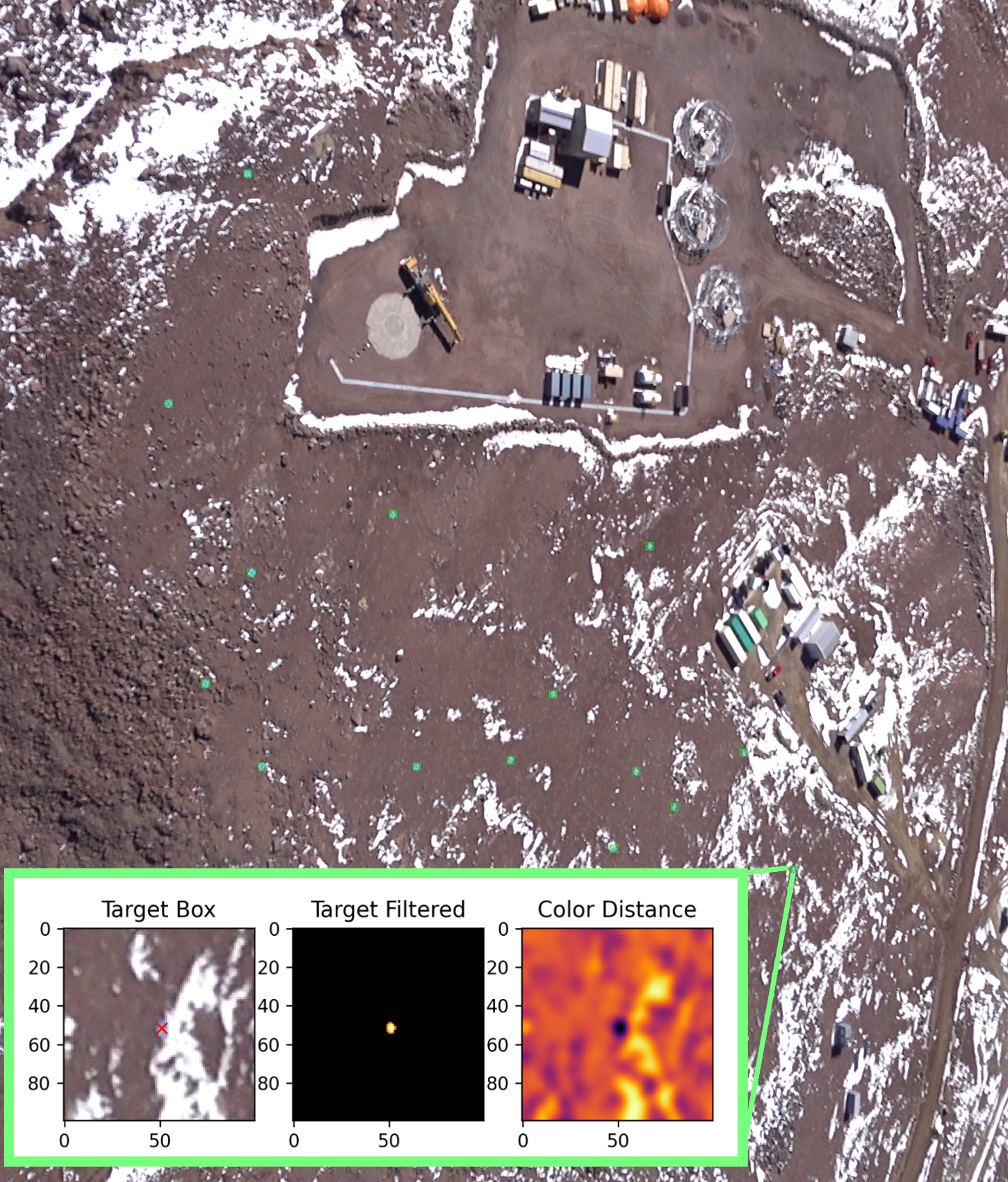}
        \caption{Single frame analysis of a 2023 flight. The targets are correctly recognized and it is shown a zoom in on a single target to demonstrate all the process performed.}
        \label{fig:targets}
\end{figure}

The three campaigns performed show clearly that the best flight strategy is to scan the telescope in azimuth and moving the drone in elevation. This particular strategy has the advantage to keep a constant atmospheric loading on the detector and we keep a constant distant of \SI{500}{\meter} with respect to the telescope chosen as the point of reference. Compared to the first campaign, where we chose to set the camera in Photo Mode with a maximum resolution of 15.3 MPx and an average frame rate of 2.5, we moved to video mode. The video resolution is set to 4k, so approximately 8 MPx, but the frame rate is constant at 30 fps. This factor of 10 increase in sampling easily justify the reduction of resolution. To compute the attitude solution, we create an algorithm that automatically recognize the targets and then compute the attitude solution using solvePnP algorithms. To detect the targets, we converted each frame to LAB color space and compute the color distance using as a metric the $\Delta E_{2000}$. The resulting distance map is then filtered to highlight the target itself so that can easily recognized. This process can be observed in Figure \ref{fig:targets}, where we highlight all the targets found and the analysis performed on a single target. Additional details on the photogrammetry system and analysis can be found in \cite{Dunner2024}.

\begin{figure}
    \centering
    \begin{minipage}{0.43\textwidth}
        \raggedleft
        \includegraphics[width=\textwidth]{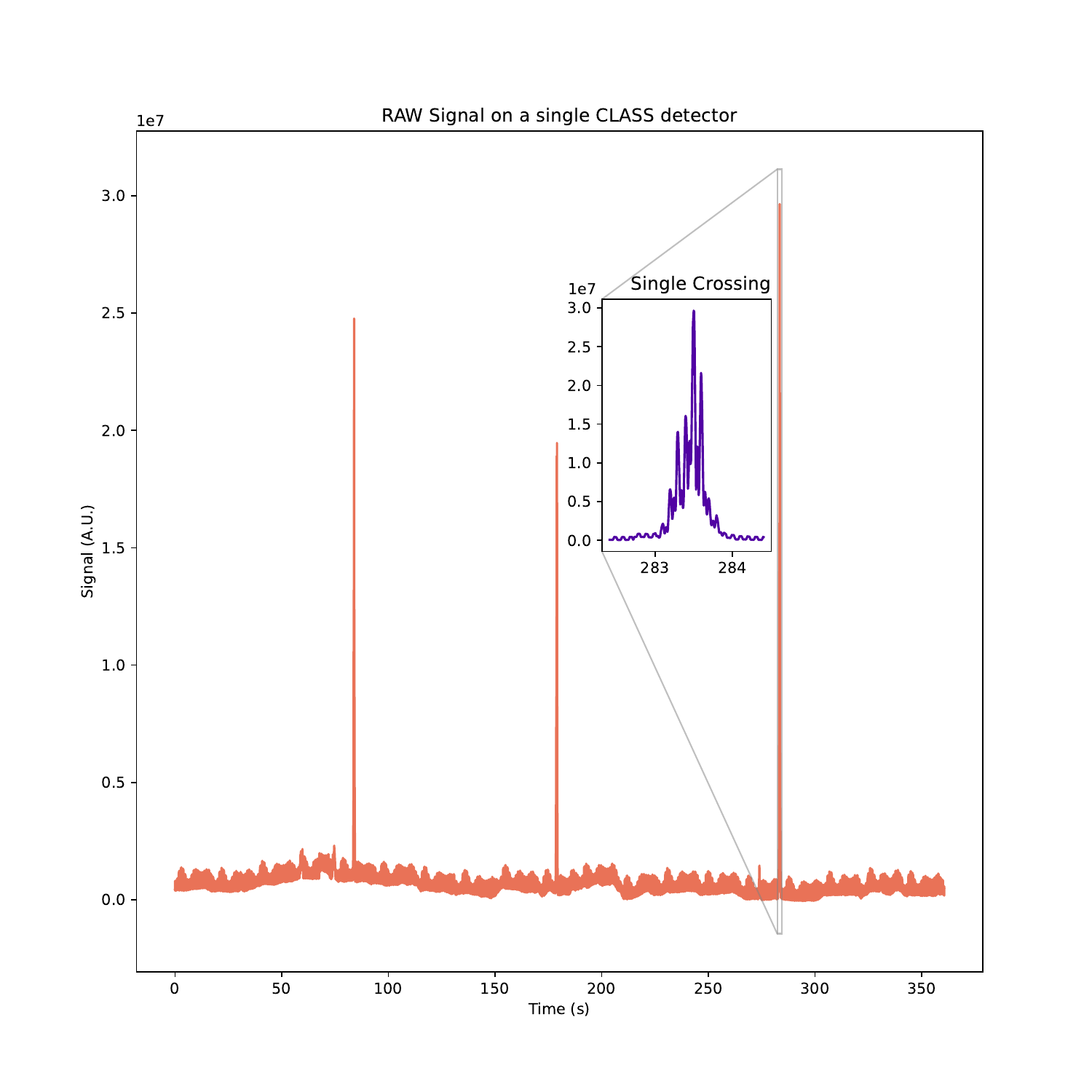}
        \caption{Shot taken by the photogrammetery camera during one flight. It is possible to notice some small with dots that represents the photogrammetry targets.}
        \label{fig:class}
    \end{minipage}
    \hspace{0.05\textwidth}
    \begin{minipage}{0.5\textwidth}
        \raggedleft
        \includegraphics[width=\textwidth]{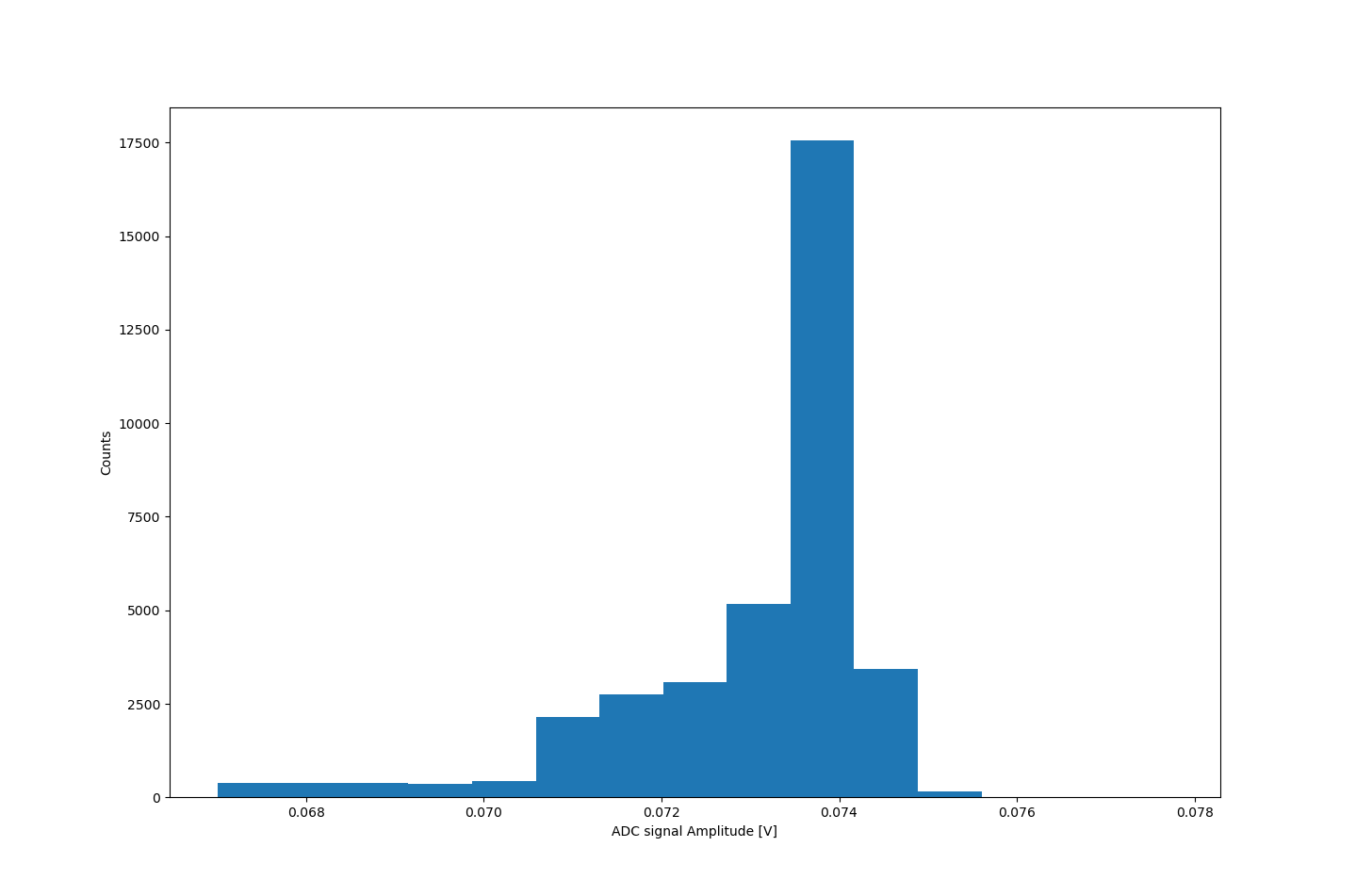}
        \caption{Histogram of the amplitude of the modulated signal throughout the flight.}
        \label{fig:adc}
    \end{minipage}
\end{figure}

Compared to the previous flight, in 2023 and 2024 we modulated the signal at \SI{47}{\hertz}. To achieve this modulation, we use the internal function of the Valon-5019, so we basically turn on and off the signal. In 2022, we found a significant thermal drift throughout the flight with the onboard flight, where the measured power was changing of $13\%$ from the beginning to the end of the flight. With the current design the thermal drift is disappered. Indeed, from Figure \ref{fig:adc}, it is possible to notice how the amplitude of each square wave measured is around the mean value of \SI{0.073}{\volt}, with just few outliers which are within $5\%$ of the mean value. The signal emitted is clearly visible by the telescopes on the ground as can be seen in Figure \ref{fig:class}, where a RAW timestream of CLASS is presenter. The data show a double modulation, the \SI{10}{\hertz} due to the VPM and the source chop at \SI{47}{\hertz}.

In 2024, we kept the same flight duration of the previous campaign to study the advantages given by the new, lighter payload and we found out that we got on average a $10\%$ more drone battery when we landed. For the next campaign, we can so increase the flight time give the additionally battery life that this payload gives.

\section{FUTURE and CONCLUSION}
\label{sec:conclusion}

In this publication, we presented the upgrades to the PROTOCALC payload and the results from the first three calibration campaigns. These were successful and show significant progress in the project development. Other than a complete payload redesign, we also open the platform to host other sources to calibrate different bands and we are ready in the next campaign to calibrate also high frequency bands of CMB telescopes. Additionally, the pointing accuracy achieved with PROTOCALC will also be increased with the use of the additional attitude sensors installed on a custom board created for this project. 

\section{Acknowledgements}
Gabriele Coppi is supported by the European Research Council under the Marie Sk\l{}odowska Curie actions through the Individual European Fellowship No. 892174 PROTOCALC. RD acknowledges support from ANID awards QUIMAL-160009, FONDEF-ID21I10236, and BASAL CATA FB210003. We thank the CLASS, SO, SA and ACT experiments for their support and contributions to this project. RD acknowledges support from ANID awards QUIMAL-160009, FONDEF-ID21I10236, and BASAL CATA FB210003. FN acknowledges funding from EU ERC POLOCALC (101096035).

\bibliography{main.bib} 
\bibliographystyle{spiebib} 

\end{document}